\documentclass[prd,preprint,nofootinbib]{revtex4-1}
\pdfoutput=1

\usepackage{amssymb,amsmath}
\usepackage{color}
\usepackage{graphicx,subfigure}
\usepackage{hyperref}
\usepackage{float} 
\usepackage[T1]{fontenc}

\begin{document}

\title{On the Statistical Treatment of the Cabibbo Angle Anomaly}

\preprint{JLAB-THY-19-3050}

\author{Yuval Grossman}
\email{yg73@cornell.edu}
\affiliation{Department of Physics, LEPP, Cornell University, Ithaca, NY 14853, USA}

\author{Emilie Passemar}
\email{epassema@indiana.edu}
\affiliation{Department of Physics, Indiana University, Bloomington, IN 47405, USA}
\affiliation{Center for Exploration of Energy and Matter, Indiana University, Bloomington, IN 47408, USA}
\affiliation{Theory Center, Thomas Jefferson National Accelerator Facility, Newport News, VA 23606, USA}

\author{Stefan Schacht}
\email{ss3843@cornell.edu}
\affiliation{Department of Physics, LEPP, Cornell University, Ithaca, NY 14853, USA \phantom{.}}

\begin{abstract}
We point out that testing the equality of the Cabibbo angle as extracted from $\Gamma(K\rightarrow \pi l\nu)$, the ratio $\Gamma(K\rightarrow l\nu)/\Gamma(\pi\rightarrow l\nu)$ and nuclear $\beta$ decays is not identical to a test of first row unitarity of the Cabibbo-Kobayashi-Maskawa (CKM) matrix. The reason is that a CKM unitarity test involves only two parameters, while 
the degrees of freedom for the assessment of the goodness-of-fit of the universality of the Cabibbo angle entailed by the Standard Model (SM) is equal to the number of measurements minus one. Beyond the SM all different processes could in principle give different Cabibbo angles. Consequently, the difference between the two tests becomes relevant starting from three observables giving results for the Cabibbo angle that are in tension with each other. With current data, depending on the treatment of the nuclear $\beta$ decays, we find that New Physics is favored over the SM at $5.1\,\sigma$ or $3.6\,\sigma$ while CKM unitarity is rejected at $4.8\sigma$ or $3.0\sigma$, respectively. We argue that the best method to test the SM is to test the equality of the Cabibbo angle, because CKM unitarity is only one aspect of the SM. 
\end{abstract}

\maketitle

\section{Introduction \label{sec:intro}}

Among several methods to determine the magnitude of the Cabibbo-Kobayashi-Maskawa (CKM) matrix elements $V_{ud}$ and $V_{us}$, the most precise ones today are the extraction from:
\begin{align}
 K_{l3} &:  \Gamma(K\rightarrow \pi l\nu), \text{where $l=\mu,e$~\cite{Sirlin:1981ie, Gasser:1984ux, Bijnens:2003uy, Cirigliano:2008wn, Antonelli:2010yf, Moulson:2017ive, Tanabashi:2018oca, Aoki:2019cca, Moulson:2019talk},} \label{eq:obsKl3}\\  
 K_{\mu 2} &:  \frac{\Gamma(K\rightarrow \mu\nu)}{\Gamma(\pi\rightarrow \mu\nu)}~\text{\cite{Antonelli:2010yf, Cirigliano:2011tm, Giusti:2017dwk, Moulson:2017ive, Tanabashi:2018oca, Aoki:2019cca, Passemar:2019talk, Kitahara:2019talk, DiCarlo:2019thl},} \label{eq:obsKl2} \\    
\beta   &: \text{Nuclear $0^+\rightarrow 0^+$ $\beta$ decays \cite{Marciano:2005ec, Hardy:2008gy, Towner:2010zz, Hardy:2014qxa, Seng:2018yzq, Seng:2018qru, Gorchtein:2018fxl, Tanabashi:2018oca, Czarnecki:2019mwq, Aoki:2019cca}.}  \label{eq:obsbeta}  
\end{align}
For brevity, we use $V_{ij}$ to denote the magnitude of a CKM matrix element.
Because of the smallness of $V_{ub}^2 \simeq 1.6\cdot 10^{-5}$~\cite{Tanabashi:2018oca} 
we can neglect $V_{ub}^2$ in the first row CKM unitarity relation, resulting in  
\begin{align}
V_{ud}^2 + V_{us}^2 &= 1\,. \label{eq:unitarity} 
\end{align}
Eq.~(\ref{eq:unitarity}) has been extensively employed in order to probe for, or constrain, new physics 
(NP)~\cite{Marciano:2004uf, Bernard:2007cf,  Antonelli:2008jg, Cirigliano:2009wk, Antonelli:2010yf, Cirigliano:2011ny, Bhattacharya:2011qm, Cirigliano:2012ab, Cirigliano:2013xha, Gonzalez-Alonso:2013uqa, Rosner:2015wva, Alioli:2017ces, Moulson:2017ive}.
Equivalently to Eq.~(\ref{eq:unitarity}), we can    
parametrize $V_{ud}$ and $V_{us}$ in the SM up to corrections of order $\mathcal{O}(\lambda^6) \simeq 0.0001$ by using the 
Cabibbo angle describing the mixing of the first two generations 
\begin{align}
V_{ud} &= \cos\theta_C\,, 	\qquad 
V_{us} = \sin\theta_C\,,
\end{align}
i.e. we practically have a two-generational model. 
A high-order Wolfenstein expansion in the Wolfenstein parameter $\lambda$ can be found in Refs.~\cite{Branco:1988ba, Charles:2004jd}.

In order to denote the origin of an extraction of the respective CKM matrix element (or their ratio) from experimental data, we use the notation $V_{us}^{K_{l3}}$, $\left(V_{us}/V_{ud}\right)^{K_{l2}}$ and $V_{ud}^{\beta}$, respectively.
As of now, there are two anomalies: 
Firstly, there is a significant tension of $V_{us}^{K_{l3}}$, $\left(V_{us}/V_{ud}\right)^{K_{l2}}$ and $V_{ud}^{\beta}$ 
with CKM unitarity~\cite{Belfatto:2019swo}. 
Second, there is an even higher tension between $V_{us}^{K_{l3}}$ and $V_{ud}^{\beta}$~\cite{Belfatto:2019swo, Aoki:2019cca, Tan:2019yqp}.

Motivated by these developments, in this paper we discuss the statistical methodology and the differences between 
testing the SM via the goodness-of-fit of the universality of the Cabibbo angle versus testing the hypothesis of CKM unitarity, Eq.~(\ref{eq:unitarity}).

We stress that we do not discuss here \lq\lq{}the\rq\rq{} global SM test that would require a global discussion of all present anomalies. Rather, the test of the goodness-of-fit of the universality of the Cabibbo angle is a SM test focusing on one aspect of the SM. 
 
Note that in a general model beyond the SM (BSM) $n$ different processes that we use as measurements of the Cabibbo angle could result in $n$ different values, giving a perfect description of the data in any case.
The number of the degrees of freedom of the comparison of the universality of the Cabibbo angle with the data is the number of different observables described by the Cabibbo angle in the SM, minus the one parameter. 
However, tests of CKM unitarity involve only two parameters, namely $V_{us}$ and the violation of unitarity $\Delta$ (see Eq.~(\ref{eq:Vud-unitarity}) below). 
In that case we compare a 
one-parameter fit to a two-parameter fit only. No matter how many measurements are available, the degree of freedom of the CKM unitarity test is always fixed.
In the past, when only two out of three measurements in Eqs.~(\ref{eq:obsKl3})--(\ref{eq:obsbeta}) showed a tension between each other, this difference was not significant.
However, when tensions between all three measurements are present, as is the current situation, one gets sensitive to the fact that 
in general the significances for the rejection of the SM via the entailed universality of the Cabibbo angle and CKM unitarity are different. 

The point that there is more to test in the measurements of $V_{us}$ and $V_{ud}$ than CKM unitarity was made 
in specific cases before \cite{Antonelli:2008jg, Antonelli:2010yf, Gonzalez-Alonso:2016etj}. Our aim here is to generalize this observation and give 
a universal methodology for SM tests with an arbitrary number of measurements of $\theta_C$. 

We emphasize that the point in this paper is only about the methodology of testing the SM with data on $V_{us}$ and $V_{ud}$. 
We do not advocate any of the extractions of $V_{ud}$, which we use as examples, and are agnostic about the validity of the used models.
Especially, we do not claim that the SM is excluded at or beyond $5\,\sigma$. 

In Sec.~\ref{sec:general} we analyze the difference between testing the SM and CKM unitarity. 
Subsequently, in Sec.~\ref{sec:current} we present our likelihood ratio tests of the SM and CKM unitarity with current data. 
In Sec.~\ref{sec:NP} we discuss a specific NP model. In Sec.~\ref{sec:conclusions} we conclude.

\section{General Testing Formalism \label{sec:general}}

\subsection{SM test: Universality test of the Cabibbo angle \label{sec:SMtest}}

In order to test the universality of the Cabibbo angle within the SM, we assess the goodness-of-fit  
of the one-parameter null hypothesis 
\begin{align}
\theta_C &= \theta_1 = \theta_2 = \dots = \theta_n\,, \label{eq:generalnullhypothesis}
\end{align} 
for $n$ different experimental determinations of the mixing angle with different observables. 
We assume here for simplicity that measurements of the same observable by different experiments are already averaged.
Beyond the SM, the analysis of $n$ different observables could in principle result in $n$ different mixing angles
of the first two generations.

The number of degrees of freedom of the test of the goodness-of-fit is therefore always one less than the total number 
of observables. 
Consequently, we calculate the two-sided $p$-value and the 
significance $z$ of the rejection of 
the SM as (see e.g. Refs.~\cite{Demortier:2007, Wiebusch:2012en,Tanabashi:2018oca})
\begin{align}
z &= \sqrt{2} \, \mathrm{Erf}^{-1}(1 - p)\,, \qquad  
p = 1 - P_{\nu/2}(\chi^2/2)\,. \label{eq:zvalue-pvalue} 
\end{align}
Here, $P_{\nu/2}(\chi^2/2)$ is the regularized lower incomplete gamma function,  
$\nu = \nu_{\text{SM test}} = n-1$ the number of degrees of freedom and $\chi^2$ the minimal $\chi^2_{\text{SM test}}$ of the one-parameter fit of the Cabibbo angle to the data in the SM. 

We note that it is inevitable that possible fluctuations of experimental measurements enter the hypothesis test, making it necessary to utilize a high threshold before rejecting the SM. For example in such a case it is necessary to identify a realistic concrete NP model that has the ability to explain the data.

\subsection{CKM unitarity test}

In order to test CKM unitarity with $n$ observables one uses two parameters $V_{us}$ and $\Delta$, the latter of which 
is used as a measure for the deviation from unitarity. We choose to employ $\Delta$ for the parametrization of $V_{ud}$ in
the form
\begin{align}
V_{ud} &= \sqrt{1 - V_{us}^2} + \Delta \,. \label{eq:Vud-unitarity}
\end{align}
We test the null hypothesis $\Delta = 0$ against the general case including $\Delta \neq 0$, which is effectively the same as 
varying $V_{us}$ and $V_{ud}$ freely. We use the $\Delta$ notation in order to make completely clear that the two models 
that we compare are nested. 
We denote the corresponding minimal $\chi^2$ values as $\chi^2_{\text{unitary}}$ and 
$\chi^2_{\text{non-unitary}}$, respectively, and define for the CKM unitarity test 
\begin{align}
\Delta \chi^2_{\text{unitarity test}} &\equiv \chi^2_{\text{min, unitary}} - \chi^2_{\text{min, non-unitary}}\,.
\end{align}
Note that $\chi^2_{\text{min, non-unitary}}$ is not necessarily zero, so that it can in principle happen that both the SM and the non-unitary model give a bad fit of the data.

\subsection{Comparison of SM test and CKM unitarity test}

\begin{table}[t]
\begin{center}
\begin{tabular}{c|c|c|c}
\hline \hline
$n$ 				  	 & \phantom{0.1cm} 1 \phantom{0.1cm}  & \phantom{0.1cm}2  \phantom{0.1cm}	   & $\geq 3$ \\\hline 
$\chi^2_{\text{SM test}}$  	 & 0  & $\chi^2$   & $\chi^2 >\Delta \chi^2_{\text{unitarity test}} $  		\\
$\nu_{\text{SM test}}$       	  	 & 0  & 1  	   & $n-1\geq 2$  		\\
$p_{\text{SM test}}$ 		  	 & 1  & $p$   	   & $\neq p_{\text{unitarity test}}$  	\\
$z_{\text{SM test}}$  		  	 & 0  & $z$   	   & $\neq z_{\text{unitarity test}}$   \\\hline
$\Delta \chi^2_{\text{unitarity test}}$  & 0  & $\chi^2$   & $< \chi^2_{\text{SM test}} $    	\\
$\chi^2_{\text{min, unitary}}$ 		 & 0  & $\chi^2$   & $\chi^2$ 	\\
$\chi^2_{\text{min, non-unitary}}$ 	 & 0  & 0   &   $> 0$ 	\\
$\nu_{\text{unitarity test}}$       	 & 1  & 1   &  $1$  		\\
$p_{\text{unitarity test}}$ 		 & 1  & $p$ & $\neq p_{\text{SM test}}$  	\\
$z_{\text{unitarity test}}$  	         & 0  & $z$ & $\neq z_{\text{SM test}}$  	\\\hline\hline
\end{tabular}
\caption{General comparison of the Cabibbo angle universality SM test and the CKM unitarity test, showing that the test results are different starting from three observables. \label{tab:general-case}}
\end{center}
\end{table}

The null hypothesis fits of the test of the SM through Cabibbo angle universality and the CKM unitarity test are equivalent. They are both one-parameter fits 
and lead to the same~$\chi^2$\,: 
\begin{align}
\chi^2_{\text{SM test}} &= \chi^2_{\text{min, unitary}}\,. \label{eq:constrained-fits}
\end{align}
However, for the SM test we assess the goodness-of-fit of the Cabibbo angle universality hypothesis, while 
the CKM unitarity test is a comparison of the hypotheses of unitary vs. non-unitary. 
The two tests have a different number of degrees of freedom.
For the CKM unitarity test the difference of dimensionality of the two theory 
spaces that we compare is always fixed to
\begin{align}
\nu_{\text{unitarity test}} &= 1\,.
\end{align}
For the goodness-of-fit test we have  
\begin{align}
\nu_{\text{SM test}} &= n - 1\,. \label{eq:dof-SM} 
\end{align}
Furthermore, the non-unitary fit allowing for $\Delta \neq 0$ is nontrivial, resulting in 
general in $\chi^2_{\text{min, non-unitary}} \neq 0$.

Whether or not the SM test and CKM unitarity test give the same results depends on the number of observables $n$ that are taken into account, as we show in Table~\ref{tab:general-case}.
\begin{itemize}
\item $n=1$ is the trivial case where no Cabibbo angle universality test is needed or possible at all, because one observable can always be fitted by one parameter. 
	Also, no violation of unitarity can possibly be detected, so everything is in agreement equally with universality and unitarity. 

\item For $n=2$, $\chi^2_{\text{SM test}} = \Delta \chi^2_{\text{unitarity test}} = \chi^2_{\text{min, unitary}}$ and 
	$\chi^2_{\text{min, non-unitary}} = 0$ because we can always explain two measurements with two free parameters. The tests have also the same number of degrees of freedom and $z_{\text{SM test}} = z_{\text{unitarity test}}$. 

\item $n\geq 3$: In this case in general the unconstrained two-parameter CKM unitarity fit cannot explain the data perfectly anymore, i.e. $\chi^2_{\text{min, non-unitary}} > 0$ and therefore we have in general $\chi^2_{\text{SM test}} > \Delta \chi^2_{\text{unitarity test}}$. 
The point is that some patterns in the data can not be accounted for by just employing a two parameter fit without unitarity.
That means for example that this procedure does not account for the generality of all possible BSM models.
\end{itemize}

\begin{figure}[t]
 \begin{center}
  \includegraphics[width=0.7\textwidth]{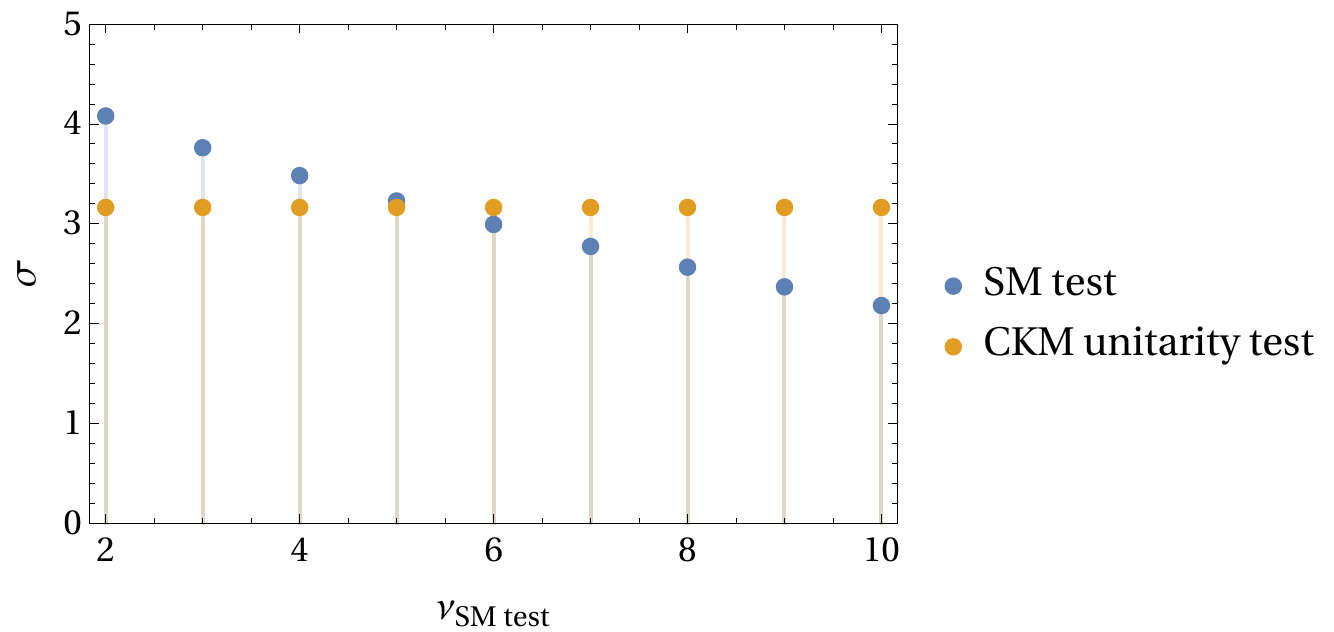} 
  \caption{Toy example for the comparison of significances of the rejection of the SM and CKM unitarity  
for fixed $\Delta \chi^2_{\text{SM test}} = 20$ and $\Delta \chi^2_{\text{unitarity test}} = 10$
as a function of $\nu_{\text{SM test}}\geq 2$. Note that $\nu_{\text{unitarity test}} = 1$ always and 
$\nu_{\text{SM test}} = n - 1$ for $n$ observables, see Eq.~(\ref{eq:dof-SM}).
Of course in reality $\Delta \chi^2_{\text{SM test}}$ and $\Delta \chi^2_{\text{unitarity test}}$ would in general also 
change when $\nu_{\text{SM test}}$ does.
However, we can see from this example that in principle either significance can be larger than the other one. 
\label{fig:zvalues}}
\end{center}
\end{figure}

Note that for $n\geq 3$ one cannot a priori say if $z_{\text{SM test}} > z_{\text{unitarity test}}$ or vice versa, 
because this does not only depend on $\chi^2_{\text{SM test}}$ and $\Delta \chi^2_{\text{unitarity test}}$ but 
also on the specific value of $\nu_{\text{SM test}} \geq 2$ vs. $\nu_{\text{unitarity test}} = 1$, that is the number of observables $n$. 
For example, for given values of $\chi^2_{\text{SM test}} = 20$ and $\Delta \chi^2_{\text{unitarity test}}=10$, we have 
$z_{\text{SM test}} > z_{\text{unitarity test}}$ if $n=5$ or $z_{\text{SM test}} < z_{\text{unitarity test}}$ if $n=8$, see  
Fig.~\ref{fig:zvalues}.

The above discussion makes clear what are the differences between unitarity tests and SM tests via universality tests of the Cabibbo angle in a completely general perspective. 
In Sec.~\ref{sec:current} we apply the above formalism to the current status of the data.

It is useful to compare our methodology to the statistical treatment of the Higgs boson~\cite{Aad:2012tfa,Chatrchyan:2012xdj}. For its discovery one compared the hypothesis of \lq{}signal\rq{} vs.~the null hypothesis of \lq{}background only\rq{}. When the null hypothesis was excluded  at $\geq 5\sigma$, we could speak of the discovery of the Higgs boson. Afterwards, different hypotheses for the properties of the Higgs boson could be tested. In our case, the \lq{}background only\rq{} hypothesis is the universality of the Cabibbo angle entailed by the SM. 
Like in the Higgs search we compare this background-only hypothesis to the signal which is observed and assess the goodness-of-fit of the background-only hypothesis by comparison to the data. The signal in our case would be the deviation of at least one of the observables from the Cabibbo angle universality hypothesis. If that was observed at $\geq 5\sigma$ the SM would be rejected.

\section{Application of formalism to current data \label{sec:current}}

Current data provides $n=3$ precision determinations of the Cabibbo angle
\begin{align}
 \sin\theta_{K_{l3}} &= V_{us}^{K_{l3}} = V_{us}         \,,  \label{eq:angle-Kl3} \\
 \cos\theta_{\beta}  &= V_{ud}^{\beta} = \left| \sqrt{1 - V_{us}^2} + \Delta\right|  \,,  \label{eq:angle-beta} \\
 \tan\theta_{K_{l2}} &= \left(\frac{V_{us}}{V_{ud}}\right)^{K_{l2}} = \frac{V_{us}}{\left|\sqrt{1 - V_{us}^2} + \Delta\right|}  \,,  \label{eq:angle-Kl2}
\end{align}
where $\theta_{K_{l3}}$, $\theta_{K_{l2}}$ and $\theta_{\beta}$ could all be different in BSM models. 
On the right hand side of Eqs.~(\ref{eq:angle-Kl3})--(\ref{eq:angle-Kl2}) we write also the expressions in terms of the parametrization for the CKM unitarity test.
In the SM, all of these extractions should be equal up to corrections of order~$\mathcal{O}(\lambda^6)$
\begin{align}
\theta_C = \theta_{K_{l3}} = \theta_{K_{l2}} = \theta_{\beta} \,. \label{eq:nullhypothesis-SM}
\end{align}
CKM unitarity on the other hand implies
\begin{align}
\Delta &= 0\,. \label{eq:nullhypothesis-CKMunitarity}
\end{align}
Eqs.~(\ref{eq:nullhypothesis-SM}) and (\ref{eq:nullhypothesis-CKMunitarity}) are the SM and CKM unitarity null hypotheses, respectively.

We summarize the latest determinations of $V_{us}$ and $V_{ud}$ in Table~\ref{tab:data}.
The obtained value for $V_{ud}$ depends on the details of the treatment of nuclear $\beta$ decays.
There are extractions available from Seng, Gorchtein, Patel, Ramsey-Musolf (SGPRM)~\cite{Seng:2018yzq, Seng:2018qru, Gorchtein:2018fxl} and Czarnecki, Marciano, Sirlin (CMS)~\cite{Czarnecki:2019mwq} using different estimates for the radiative corrections.

Our fit results are shown in Table~\ref{tab:testresults}. 
Therein, also subsets of observables are considered for illustration purposes. 
As discussed in Sec.~\ref{sec:general}, for $n=2$ fits the Cabibbo angle universality SM test and CKM unitarity test 
give the same results, and for the current full data set with $n=3$ they differ.
While the difference of significances of CKM unitarity test and Cabibbo angle universality SM test is not dramatic, in case of the SPRGM interpretation the significances of rejection of Cabibbo angle universality and CKM unitarity are $5.1\sigma$ vs. $4.8\sigma$, and for the CMS interpretation $3.6\sigma$ vs $3.0\sigma$, respectively.

\begin{table}[t]
\begin{center}
\begin{tabular}{c|c|c|c}
\hline \hline
Observable & Measurement & Method  & References  \\\hline
$\vert V_{us}\vert^{K_{l3}}$ & $0.22326\pm0.00058$  & $K_{l3}$ decays 	  &   \cite{Antonelli:2010yf,Moulson:2019talk} \\
$\vert V_{us}/V_{ud}\vert^{K_{l2}}$ &  $0.23129\pm 0.00045$ & $K_{l2}/\pi_{l2}$ decays & \cite{Antonelli:2010yf,Passemar:2019talk, Kitahara:2019talk} \\
$\vert V_{ud}\vert^{\beta}$ & $0.97370\pm 0.00014$ & Nuclear $\beta$ decays, SGPRM extraction & \cite{Seng:2018yzq, Seng:2018qru, Gorchtein:2018fxl} \\
$\vert V_{ud}\vert^{\beta}$ & $0.97389\pm 0.00018$ & Nuclear $\beta$ decays, CMS extraction & \cite{Czarnecki:2019mwq} \\\hline\hline
\end{tabular}
\caption{Observables and data used in the fits. In case of the new physics scenario these are interpreted as effective values, see Eqs.~(\ref{eq:NP-Kl3})--(\ref{eq:NP-beta}). 
$V_{us}$ and $V_{us}/V_{ud}$ have been extracted from kaon decays~\cite{Antonelli:2010yf,Moulson:2019talk,Passemar:2019talk, Kitahara:2019talk}
using the $N_f = 2+1+1$ lattice results~\cite{Aoki:2019cca,Bazavov:2018kjg}.
The obtained value for $V_{ud}$ depends on the details of the treatment of nuclear $\beta$ decays.
There are extractions available from Seng, Gorchtein, Patel, Ramsey-Musolf (SGPRM) \cite{Seng:2018yzq, Seng:2018qru, Gorchtein:2018fxl} and 
Czarnecki, Marciano, Sirlin (CMS)~\cite{Czarnecki:2019mwq} using different estimates for the radiative corrections.
\label{tab:data}}
\end{center}
\end{table}

\begin{table}[t]
\begin{center}
\begin{tabular}{c|c|c|c|c|c|c|c|c}
\hline \hline

Fit & $n$ & $\chi^2_{\text{SM test}}$  	& $\nu_{\text{SM test}}$    &  $p_{\text{SM test}}$ &  $z_{\text{SM test}}$  &  $\Delta \chi^2_{\text{unitarity test}}$  &  $p_{\text{unitarity test}}$ &  $z_{\text{unitarity test}}$  \\\hline

$K_{l3}+K_{l2}$		    & 2 & 8.5  & 1	& 0.0036  & 2.9 $\sigma$ & 8.5	&   0.0036	& 2.9 $\sigma$	 \\\hline 

$K_{l3}+K_{l2}+\beta$ (SGPRM) & 3 & 30.0	& 2	& $3.1\cdot 10^{-7}$  & 5.1 $\sigma$ & 22.8	&  $1.8\cdot 10^{-6}$ & 4.8 $\sigma$  \\ 

$K_{l2}+\beta$ (SGPRM) 	    & 2 & 11.6	& 1	& 0.00065 & 3.4 $\sigma$ & 11.6	&   0.00065 & 3.4 $\sigma$	\\ 

$K_{l3}+\beta$ (SGPRM) 	    & 2 & 30.0 & 1	& $4.4\cdot 10^{-8}$	&   5.5 $\sigma$ & 30.0	&  $4.4\cdot 10^{-8}$	& 5.5 $\sigma$	  \\\hline
 
$K_{l3}+K_{l2}+\beta$ (CMS) & 3 & 16.5	& 2	& 0.00027 & 3.6  $\sigma$   & 9.0	&  0.0027 & 3.0 $\sigma$ 	   	      \\ 

$K_{l2}+\beta$ (CMS)        & 2 & 3.6 & 1	& 0.056	& 1.9  $\sigma$   & 3.6	&  0.056 	&  1.9 $\sigma$  	      \\ 

$K_{l3}+\beta$ (CMS) 	    & 2 & 15.1 & 1     &  0.00010    &  3.9  $\sigma$  & 15.1	&  0.00010 & 3.9 $\sigma$      \\\hline\hline 
  
\end{tabular}
\caption{Cabibbo angle universality SM test and CKM unitarity tests for different data sets. $z_{\text{SM test}}$ is the significance of the rejection of Cabibbo angle universality and $z_{\text{unitarity test}}$ is the significance of the CKM unitarity rejection. \label{tab:testresults}}
\end{center}
\end{table}

\section{New Physics Models \label{sec:NP}}

In this section, which is based on an idea first put forward in Refs.~\cite{Passemar:2019talk, Kitahara:2019talk}, we demonstrate the ability of a concrete BSM model to describe the data with $\chi^2_{\text{min, BSM}}=0$, 
while pointing out that it is not even clear how to formulate the corresponding fit in terms of a CKM unitarity test.
We emphasize that this serves as a toy example for illustration only, that is, we did not apply all the available constraints. 
 
We employ the model and notation of Ref.~\cite{Bernard:2007cf} and show that BSM 
couplings of right-handed (RH) quarks \cite{Pati:1974vw, Mohapatra:1974gc, Mohapatra:1974hk, Wilczek:1975kn, Senjanovic:1975rk, Senjanovic:1978ev, Bernard:2006gy, Towner:2010zz, Senjanovic:2014pva, Cirigliano:2016yhc, Alioli:2017ces} to the $W$ boson, i.e.~RH currents, could remove the tensions presented in Table~\ref{tab:testresults}. 

The above model serves only as an example. We are aware that models with sterile neutrinos~\cite{Lee:1977tib, Bryman:2019bjg} may have similar effects on the CKM extraction.
Further BSM studies, which also explore the connection of kaon and $\beta$ decays to lepton flavor non-universality can be found in Refs.~\cite{Coutinho:2019aiy, Crivellin:2020lzu}.

Following the notation of Ref.~\cite{Bernard:2007cf}, we denote the respective coupling of RH strange quarks by
$\varepsilon_s$ and the one of down quarks by $\varepsilon_{ns}$. 
Furthermore, the measured values of the CKM matrix elements given in Table~\ref{tab:data} are 
interpreted as effective ones and are related to the mixing angle and the RH couplings as~\cite{Bernard:2007cf} 
\begin{align}
V_{us}^{K_{l3}} &= 
	\left|\sin\theta_C + \varepsilon_{s} \right| \,, \label{eq:NP-Kl3}\\
\left( \frac{V_{us}}{V_{ud}}\right)^{K_{l2}} &= 
	\left|\frac{\sin\theta_C - \varepsilon_{s}}{\cos\theta_C - \varepsilon_{ns}}\right|\,, \label{eq:NP-Kl2}\\
V_{ud}^{\beta} &= \left|\cos\theta_C + \varepsilon_{ns}\right|\,. \label{eq:NP-beta} 
\end{align}
Note that $\varepsilon_{s}$ and $\varepsilon_{ns}$ are in general complex. 
However, to keep things simple for our purposes it is enough to study the real case here.
The SM is obtained in the limit 
\begin{align}
 \varepsilon_{s} = \varepsilon_{ns} = 0\,. 
\end{align}
Considering Eqs.~(\ref{eq:NP-Kl3})--(\ref{eq:NP-beta}) it is not clear how one could rephrase this parametrization in order to perform a CKM unitarity test. 

Fitting the general model of right handed currents Eqs.~(\ref{eq:NP-Kl3})--(\ref{eq:NP-beta}),  
we obtain a perfect description of the data with $\chi^2_{\text{min,RH}}=0$.
Moving to a different model, in case we switch off the down-quark right handed currents $\varepsilon_{ns}=0$ we have a more constrained fit. We perform a likelihood ratio test 
comparing only strange RH currents with the more general case of strange and down RH currents and define 
\begin{align}
\Delta \chi^2 &\equiv \chi^2_{\text{min,RH strange}} -  \chi^2_{\text{min,RH}}\,.
\end{align}

We consider only toy NP fits to the SGPRM data set as only for that data set there is a tension with the universality of the Cabibbo angle beyond $5\,\sigma$, 
and compare the toy model with RH strange quark currents to a more general toy model that includes both strange and down quark RH currents. The relatively fixed number of parameters is always one. For any two observables out of Eqs.~(\ref{eq:obsKl3})--(\ref{eq:obsbeta}), we obtain a vanishing $\Delta \chi^2$. 
However, once we take all observables Eqs.~(\ref{eq:obsKl3})--(\ref{eq:obsbeta}) into account, we get $\Delta \chi^2 = 25.2$ and a significance of rejection of $z=5.0\,\sigma$. 
This example makes it completely obvious that it is very important to include all available data for any test for NP.

While the CKM unitarity test is a smoking gun for the presence of new physics, it is not clear how to relate it to the considered model with RH currents. The above procedure on the other hand is completely unambiguous. Furthermore, the CKM unitarity test is included in the SM test as outlined in Sec.~\ref{sec:SMtest}. 
Both tests are however of course subject to the caveat of possible statistical fluctuations. In general, by using relations between the several Cabibbo angle values the SM test can be transformed to a test of an arbitrary NP model, while the unitarity test applies only to a subset of NP models that can be mapped on a two-parameter fit.

\section{Conclusions \label{sec:conclusions}}

Recent precision determinations of $V_{us}$ and $V_{ud}$ enable unprecedented tests of the SM and 
constraints on possible NP models like right-handed currents.
We showed that a SM test via the test of the Cabibbo angle universality goes beyond just a test of CKM unitarity and 
gives different test results if more than two observables are taken into account.

In a CKM unitarity test one compares a constrained fit with a fit of free floating $V_{us}$ and $V_{ud}$.
The latter can not necessarily describe the data as well as a BSM model, in case the patterns go beyond just violating unitarity.
This matters starting from three independent observables being taken into account.
In a SM test on the other hand one assesses the goodness-of-fit of a universal Cabibbo angle by comparison to the data.
This gives the same $\chi^2$ as the CKM unitarity fit, however 
$\chi^2_{\text{SM test}}\neq \Delta \chi^2_{\text{unitarity test}}$. 
We demonstrated explicitly that in a concrete BSM model all measured effective angles could in principle be different.

Altogether, that means that the significance of SM tests can in general be different from the one of CKM unitarity fits once more than two observables are considered.  
In the foreseeable future, $\tau$~decays may provide a further precision determination of the Cabibbo angle via the ratio 
$\Gamma(\tau \rightarrow K \nu_{\tau}) / \Gamma(\tau \rightarrow \pi \nu_{\tau})$, see Refs.~\cite{Pich:2013lsa, Lusiani:2018zvr}, and the number of precision observables for the determination of the Cabibbo angle rises to four. 
Further input is also coming up from pion beta decays~\cite{Czarnecki:2019iwz}.
With more measurements in the future the differences between Cabibbo angle universality SM tests and CKM unitarity tests could become even more significant.

Consequently, we encourage to test the SM by testing for the universality of the Cabibbo angle, rather than testing for CKM unitarity only, with the general methodology laid out above. 

\begin{acknowledgments}
We thank Andreas Crivellin, Martin Hoferichter and Teppei Kitahara for discussions.
The work of YG is supported in part by the NSF grant PHY1316222.
SS is supported by a DFG Forschungs\-stipendium under contract no. SCHA 2125/1-1.
This work is supported in part by the U.S. Department of
Energy (contract DE-AC05-06OR23177) and National Science Foundation (PHY-1714253).
\end{acknowledgments}

\bibliography{draft.bib}
\bibliographystyle{apsrev4-1}

\end{document}